\documentclass[12pt]{article}
\usepackage{epsf}
\usepackage{graphicx}
\usepackage{amsfonts}
\usepackage{amsmath}
\usepackage[mathscr]{eucal}
\usepackage{booktabs}

\def\href#1#2{#2}   
\newif\ifdraft
\draftfalse	  
\reversemarginpar   
\makeatletter
\let\mlabel=\label
\let\adkendequation=\endequation%
\def\endequation{\adkendequation\adklabel\global\@ignoretrue}
\let\adkendeqnarray=\endeqnarray%
\def\endeqnarray{\adkendeqnarray\adklabel\global\@ignoretrue}
\newbox\marglabbox
\def\adklabel{\ifvoid\marglabbox\else\marginpar{\unhbox\marglabbox}\fi}
\def\label#1{\ifdraft\ifmmode%
  \global\setbox\marglabbox=\hbox{\hfill\fbox{\tiny\verb*~#1~}}%
  \else\ifinner\else\marginpar{\hfill\fbox{\tiny\verb*~#1~}}%
  \fi\fi\fi \mlabel{#1}}
\makeatother
\ifdraft%
\fi
\font\twelvefrak=eufm10 scaled 1200
\font\tenfrak=eufm10
  \newfam\frakfam
  
  \textfont\frakfam=\twelvefrak
  \scriptfont\frakfam=\tenfrak
  \scriptscriptfont\frakfam=\scriptfont\frakfam
\def\sqr#1#2{{\vcenter{\hrule height.#2pt
   \hbox{\vrule width.#2pt height#1pt \kern#1pt
      \vrule width.#2pt}
   \hrule height.#2pt}}}

\def\bsqr#1#2{{\vrule width #1pt height#2pt}}
\def\bsquare{{\mathchoice\bsqr66\bsqr66\bsqr33\bsqr33}}
\def\badbreak{\penalty1000}

\def\Trs{\mathop{\rm tr}}		    
\newcommand{\cP}{{\cal P}}                  
\newcommand{\psibar}{{\bar\psi}}            
\def\plp{{\Gamma}}                          
\def\cop{{C}}                               
\def\chps{{\Lambda_{ch}}}                   
\def\pr{{\cP}}                              
\begin{document}

\begin{center}
{\Large{\bf Spontaneous Chiral Symmetry Breaking as}} \\
\vspace*{.14in}
{\Large{\bf Condensation of Dynamical Chirality}} \\
\vspace*{.24in}
{\large{Andrei Alexandru$^1$ and Ivan Horv\'ath$^2$}}\\
\vspace*{.24in}
$^1$The George Washington University, Washington, DC, USA\\
$^2$University of Kentucky, Lexington, KY, USA

\vspace*{0.13in}
{\large{Nov 14 2012}}

\end{center}

\vspace*{-0.08in}

\begin{abstract}

\noindent
The occurrence of spontaneous chiral symmetry breaking (SChSB) is 
equivalent to sufficient abundance of Dirac near--zeromodes. However, 
dynamical mechanism leading to breakdown of chiral symmetry should be 
naturally reflected in {\em chiral properties} of the modes. Here we 
offer such connection, presenting evidence that SChSB in QCD proceeds 
via the appearance of modes exhibiting dynamical tendency for local 
chiral polarization. These modes form a band of finite width 
$\chps$ ({\em chiral polarization scale}) around the surface of otherwise 
anti--polarized Dirac sea, and condense. $\chps$ characterizes the
dynamics of the breaking phenomenon and can be converted to a quark 
mass scale, thus offering conceptual means to determine which quarks 
of nature are governed by broken chiral dynamics. 
It is proposed that, within the context of SU(3) gauge theories with 
fundamental Dirac quarks, mode condensation is equivalent to chiral 
polarization. This makes $\chps$ an ``order parameter'' of SChSB, albeit
without local dynamical field representation away from chiral limit. 
Several uses of these features, both at zero and finite temperature, 
are discussed. Our initial estimates are 
$\chps \approx 150$ MeV ($N_f\!=\!0$), $\chps \approx 80$ MeV 
($N_f\!=\!2\!+\!1$, physical point), and that the strange quark 
is too heavy to be crucially influenced by broken chiral symmetry.

\end{abstract}

\vspace*{0.10in}

\noindent{\bf 1. Introduction and Conclusions.}
Chiral symmetry and its conjectured dynamical breaking pattern in multi--flavor
massless QCD is a crucial ingredient in the current understanding of low 
energy hadronic physics. While SChSB scenario is widely accepted, and there 
is no first--principles evidence suggesting otherwise, it is not known how strong 
interactions induce its vacuum to become a non--symmetric state. To help 
the identification of the corresponding mechanism, it is desirable to search 
for dynamical circumstances accompanying the phenomenon, i.e. to find
dynamical features that the eventual explanation needs to incorporate.
Low--lying Dirac modes are a suitable place to look for such signatures since 
they encode the nature of quark propagation in the chiral regime as well
as the condensate itself. Indeed, the most direct expression of this connection
is the Banks--Casher relation, revealing that the symmetry breakdown is 
equivalent to sufficient accumulation of Dirac near--zeromodes~\cite{Ban80A}.
However, being entirely generic, the Banks--Casher relation doesn't shed 
light on dynamical specifics of SChSB. 

The premise of this work is that dynamical features of the theory relevant 
to chiral symmetry breaking should be imprinted in {\em chiral properties} 
of Dirac eigenmodes. While there cannot be any average preference for left 
or right, QCD dynamics induces specific chiral properties in the eigenmodes 
locally. The most basic of these describe whether values 
$\psi(x)=\psi_L(x)+\psi_R(x)$ tend to involve asymmetric participation 
of left-right subspaces (chiral polarization) 
as opposed to equal participation (chiral anti--polarization). 
Building on the earlier local chirality approach of Ref.~\cite{Hor01A}, 
proper {\em dynamical} quantifiers of this type have recently been 
constructed~\cite{Ale10A}.

Dynamical nature of these new polarization measures stems from the fact that 
they are defined relative to the case of statistically independent left--right 
components, and thus represent (uniquely constructed) correlations. 
In this work we will only be concerned with the overall dynamical tendency 
described by the correlation coefficient of chiral polarization 
$\cop_A$. Let $\pr (\psi_L,\psi_R)$ be the probability distribution defined 
by the collection of values comprising given mode(s), and 
$\pr^u (\psi_L,\psi_R)$ the associated distribution of statistically 
independent components. If $\plp_A$ is the probability that a sample chosen 
from $\pr$ is more polarized than the sample chosen from $\pr^u$, then 
$\cop_A \equiv 2 \plp_A -1 \in [-1,1]$. Thus, the enhancement of polarization 
relative to statistical independence ($\plp_A > 1/2$) is associated with 
correlation ($\cop_A > 0$) while its suppression ($\plp_A < 1/2$) with 
anti--correlation ($\cop_A < 0$).

\begin{figure}[t]
\begin{center}
    \centerline{
    \hskip 0.00in
    \includegraphics[width=18.0truecm,angle=0]{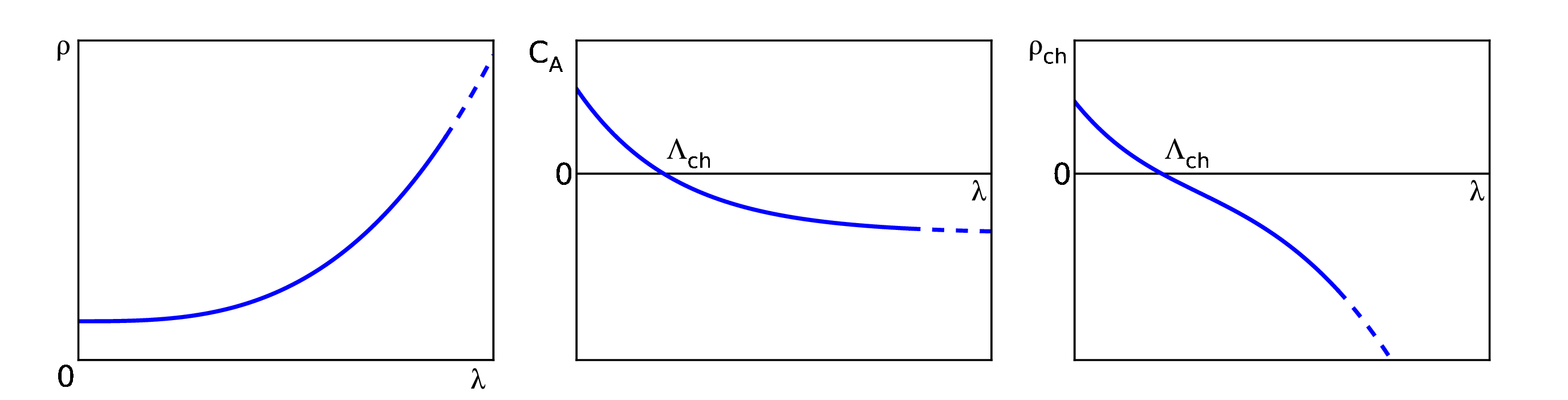}
     }
     \vskip -0.10in
     \caption{Schematic behavior of $\rho(\lambda)$, $\cop_A(\lambda)$, $\rho_{ch}(\lambda)$ 
     in theory with Dirac mode condensation.}
     \label{fig:ilustracia}
    \vskip -0.45in
\end{center}
\end{figure} 

The claims in this work are mostly based on the proposed behavior of 
the QCD average for $\cop_A$ in modes at eigenvalue $\lambda$. 
In particular, for quark setups relevant to real world, such as $N_f=2+1$
at zero temperature, we conclude the behavior shown in 
Fig.~\ref{fig:ilustracia} (middle) at generic quark masses: there is 
a low--lying band of chirally polarized modes 
separated from the rest of anti--polarized bulk. The suggestion is that this 
applies in the infinite volume and also when light quark masses are 
asymptotically small, with {\em chiral polarization scale} $\chps$, introduced
in Ref.~\cite{Ale10A}, remaining strictly positive.\footnote{Chiral polarization 
scale was denoted $\Lambda_T$ in Ref.~\cite{Ale10A}. Here we switch
to $\chps$ so that the label is not confused with temperature.} 
Since broken chiral dynamics of quarks is dominated by near--zeromodes, 
we are proposing that this dynamics (1) is associated with chirally polarized 
(correlated) modes, and that (2) the mechanism generating the polarized band 
involves a dynamical scale $\chps$. These definite properties need to be 
respected by viable models of SChSB, 
exemplifying a ``bottom--up'' approach to QCD vacuum structure~\cite{Hor06A}. 

Thus, our {\em Conjecture 1} implies that the chirality--related dynamical 
feature of SChSB in QCD is that its condensing modes are locally polarized. 
How important is this in the context of the mechanism generating broken chiral 
dynamics? One way to approach this is to consider a wider range of theories,
governed by the same gauge interaction, and ask whether the presence of broken 
chiral dynamics is equivalent to chiral polarization, i.e. whether SChSB occurs 
only in conjunction with chiral polarization and vice versa. Indeed, if such 
correspondence holds, then chiral polarization is intimately tied to the nature 
of the interaction. For purposes of this discussion, we will use the context 
of SU(3) gauge theories with arbitrary number N$_f$ of Dirac fermions in 
the fundamental representation, and at arbitrary temperatures. Our 
{\em Conjecture 2} implies that the above equivalence holds.

Accepting the relevance of chiral polarization elevates $\chps$ into a scale 
associated with SChSB. This input can be used to define the scale of SChSB in 
a more conventional language of quark mass. Resulting chiral polarization mass 
scale $m_{ch}$ represents maximal valence mass at which spectral components 
of the associated scalar bilinear favor chiral polarization over chiral 
anti--polarization on average. From chiral standpoint, quark dynamics at lower 
mass is similar to massless one, while that at higher mass turns qualitatively 
different. Moreover, $\chps$ and $m_{ch}$ are meaningful even when all quarks 
are massive, and some of them can still be driven by broken--like dynamics if their 
mass is lower than theory's $m_{ch}$. Thus, one utility of the above dynamical 
insight is a possibility of objective labeling nature's 
quarks as ``light'' or ``heavy''. These dynamical features also put a novel 
angle on the characterization of strong dynamics at finite temperature, 
as we discuss in some detail.

The proposed connection between SChSB and dynamical chirality is associated 
with the following narrative. Viewing Dirac eigenmodes as scale--dependent 
probes of gauge field, their non--interacting baseline is a perfectly 
anti--polarized state. Indeed, left and right components of free eigenmodes 
have identical magnitudes, independently of $\lambda$ in this scale--invariant 
situation, and there is no mode condensation. Turning on the interaction among 
gluons (pure glue QCD) works against anti--polarization. This effect is scale 
dependent, due to running coupling, with modes in the infrared being affected 
more than those in the ultraviolet. At low energy the interaction becomes 
sufficiently strong for the chiral behavior of modes to undergo a qualitative 
change: scale $\chps$ is dynamically generated via appearance of chirally 
polarized, condensing Dirac modes, while logarithmically violated scale 
invariance remains in the ultraviolet. The above two cases (free and pure glue) 
represent chiral extremes, with other gauge--quark setups interpolating between 
them. Indeed, both light dynamical quarks and the temperature tend 
to reduce the effective gauge coupling at low energy, and thus possibly destroy 
chiral polarization relative to pure glue. Our conclusion in most general form 
is that the Dirac mode condensation and chiral polarization are present/absent 
simultaneously: chiral polarization scale $\chps$ is an indicator of mode
condensation and, when some of the quarks are massless, an ``order parameter'' 
of SChSB.

\medskip

\noindent{\bf 2. Definitions and Claims.}
Here we specify our main conclusions more precisely. The setup and notation of 
continuum Euclidean QCD is used for simplicity, but it should be understood that 
the concepts below have well--defined meaning acquired via lattice regularization 
respecting chirality, e.g. lattice QCD with overlap fermions~\cite{Neu98BA}. 
Thus, the spectrum of continuum massless Dirac operator in a given gauge background 
is located on imaginary axis, and the zero--temperature theory with N$_f$ flavors 
of quarks can be labeled by masses $M \equiv (m_1,m_2,\ldots,m_{N_f})$. Ordering 
the eigenmodes $\psi_k(x)$ by magnitude of their eigenvalue $i\lambda_k$, 
the average correlation of chiral polarization for modes at $i\lambda$ reads
\begin{equation}
   \cop_A(\lambda,M,V) \equiv 
   \frac{\sum\limits_k \langle \, \delta(\lambda - \lambda_k) \, \cop_{A,k} \,\rangle_{M,V}}
        {\sum\limits_k \langle \, \delta(\lambda - \lambda_k) \, \rangle_{M,V}}
   \,=\,
   \frac{\rho_{ch}(\lambda,M,V)}{\rho(\lambda,M,V)}
    \label{eq:101}
\end{equation}
Here $\langle \ldots \rangle_{M,V}$ denotes QCD expectation value in 4--volume $V$, 
$\cop_{A,k}$ is the correlation associated with mode $\psi_k$, and we have introduced 
the spectral chiral polarization density 
\begin{equation}
  \rho_{ch}(\lambda,M,V) \equiv \frac{1}{V} \, 
  \sum\limits_k \langle \, \delta(\lambda - \lambda_k) \, \cop_{A,k} \,\rangle_{M,V}
  \label{eq:111}
\end{equation} 
in addition to spectral mode density $\rho(\lambda,M,V)$. Correlation $\cop_{A,k}$ 
can be evaluated relative to mode's own distribution $\pr^u_k$ of independent 
left--right components, or relative to $\pr^u_\lambda$ involving all modes at eigenvalue 
$i\lambda$. These definitions are expected to be equivalent in the infinite volume limit.
To include setups at finite temperature $T$ one simply replaces labels 
$V \rightarrow T,V_3$ and factors $1/V \rightarrow T/V_3$, where $V_3$ is a 3--volume. 

There are two reasons for introducing $\rho_{ch}$. First, spectral contribution to 
fermionic bilinears is proportional to spectral mode density, and thus 
$\rho \hspace{1pt} \cop_A = \rho_{ch}$ gives the proper spectral weight to local 
chirality in that context. Secondly, while $\cop_A(\lambda)$ is not defined in 
potential regions of spectral exclusion, i.e. parts of real axis with no 
eigenvalues in any background, $\rho_{ch}(\lambda)$ assumes definite value for each 
$\lambda$. Note that, when $\rho(\lambda)$ is non--zero, $\cop_A(\lambda)$ and 
$\rho_{ch}(\lambda)$ have identical signs and zeroes, as seen e.g. in 
Fig.~\ref{fig:ilustracia} (right).

Theory is said to exhibit ``mode condensation'' if
$\lim_{\lambda \to 0} \lim_{V\to \infty} \rho(\lambda,M,V) \!\equiv\! \rho(0,M) \!>\!0$, 
with obvious generalization to finite temperature. Similarly, ``chirality condensation''
occurs when $\rho_{ch}(0,M) > 0$, and ``anti--chirality condensation'' when
$\rho_{ch}(0,M)<0$. By ``QCD'' we mean SU(3) gauge theories in three spatial dimensions 
with Dirac quarks in the fundamental representation. 
The statements below are formulated in infinite volume where the concepts of spontaneous 
symmetry breaking and condensation assume their meaning. 

\medskip
\noindent {\bf Conjecture 1:}$\;$ 
{\sl Consider zero--temperature QCD with N$_f$=2+1 flavors of quarks, i.e. 
$M=(m_l,m_l,m_h)$, $\,m_l \in [0,\infty]$, $\, m_l < m_h$. For every $M$ there exists 
$\chps \equiv \chps(M) > 0$ such that functions $\rho_{ch}(\lambda)$ and $\cop_A(\lambda)$ 
are positive on $[0,\chps)$ and negative on $(\chps,\infty)$.}
\medskip

\noindent 
Note that the above statement subsumes the standard expectation that Dirac modes
in N$_f$=2+1 QCD always condense. Indeed, since $|\rho_{ch}(\lambda)| \le \rho(\lambda)$,
it follows that $\rho(0)>0$ for all $M$. The novelty is in the claim that this
condensation is always associated with chirally polarized modes and non--zero $\chps$, 
including in the massless ($m_l=0$) limit. The latter makes this behavior a dynamical feature 
of SChSB in QCD. Similar conclusion is expected to hold in other physically relevant cases 
such as N$_f$=2 or N$_f$=2+1+1+1+1.

\medskip
\noindent {\bf Conjecture 2:}$\;$ 
{\sl Consider QCD theories with arbitrary N$_f \ge 0$ quark flavors of arbitrary 
masses $M$, at arbitrary temperature $T$. Within this set,
$\rho(0,M,T) > 0$ if and only if there exists $\chps \equiv \chps(M,T) > 0$
such that $\rho_{ch}(\lambda)$ and $\cop_A(\lambda)$ are positive on $[0,\chps)$ 
and negative on $(\chps,\infty)$. Moreover, $\rho(0,M,T)=0$ if and only if
$\rho_{ch}(\lambda)\le 0$ on $[0,\infty)$.}
\medskip

\noindent We thus propose that, in the context of SU(3) gauge theories with fundamental 
quarks, Dirac mode condensation is synonymous with chiral polarization in the above 
sense, including in chiral corners of the theory space. The statement implies that 
there are two possible behaviors of $\rho_{ch}(\lambda)$: either 
there is anti--polarization over the whole spectrum or there exists a finite layer of 
polarization around the surface of otherwise anti--polarized Dirac sea. The associated 
two classes of theories correspond to those defined by absence/presence of Dirac mode 
condensation. We wish to highlight the following corollary.

\medskip
\noindent {\bf Corollary 1:}$\;$ 
{\sl Condensation of Dirac modes in QCD theories is a condensation of 
dynamical chirality, and so is SChSB. Anti--chirality doesn't condense.}
\medskip

We emphasize that the above statements should be understood as being valid in
lattice--regularized theory at sufficiently large cutoffs, with appropriate limiting
procedures implied. The precise value of chiral polarization scale $\chps$ for given 
continuum theory likely depends on the details of this regularization, and finite 
normalization procedure might be needed to make it unique. However, the properties 
discussed here only depend on $\chps$ being positive, and are thus expected 
to be universal. 

\begin{figure}[t]
\begin{center}
    \centerline{
    \hskip 0.08in
    \includegraphics[width=8.7truecm,angle=0]{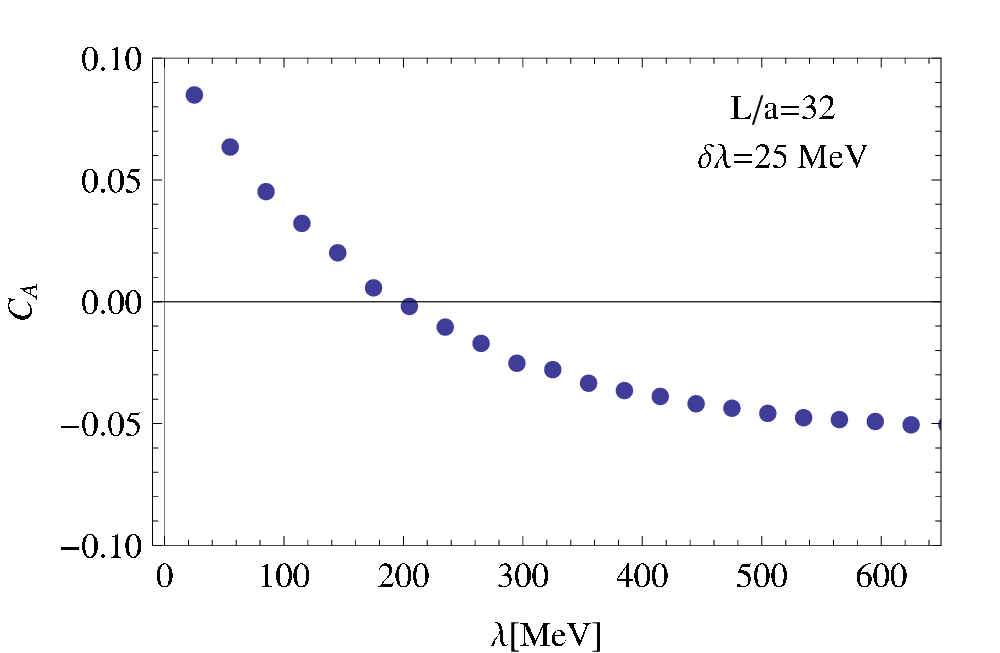}
    \hskip -0.1in
    \includegraphics[width=8.7truecm,angle=0]{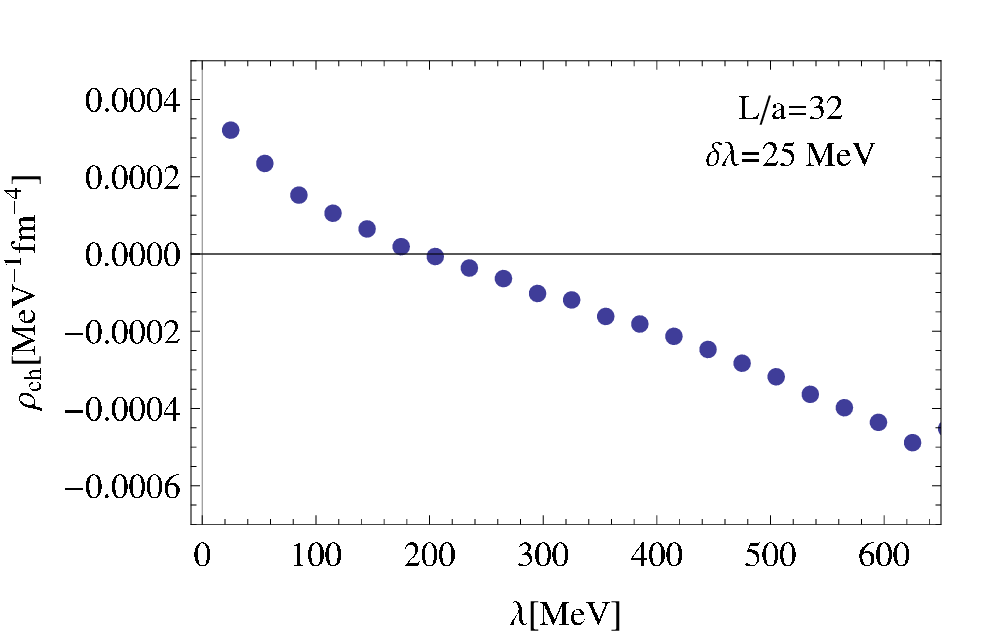}
     }
    \vskip -0.02in
    \centerline{
    \hskip 0.08in
    \includegraphics[width=8.7truecm,angle=0]{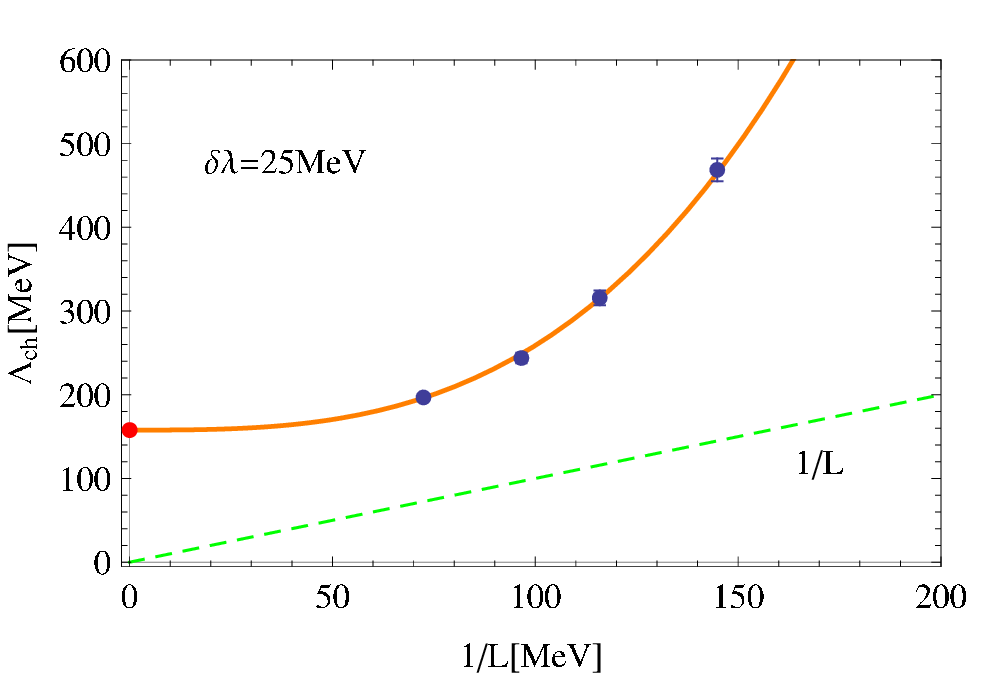}
    \hskip -0.1in
    \includegraphics[width=8.7truecm,angle=0]{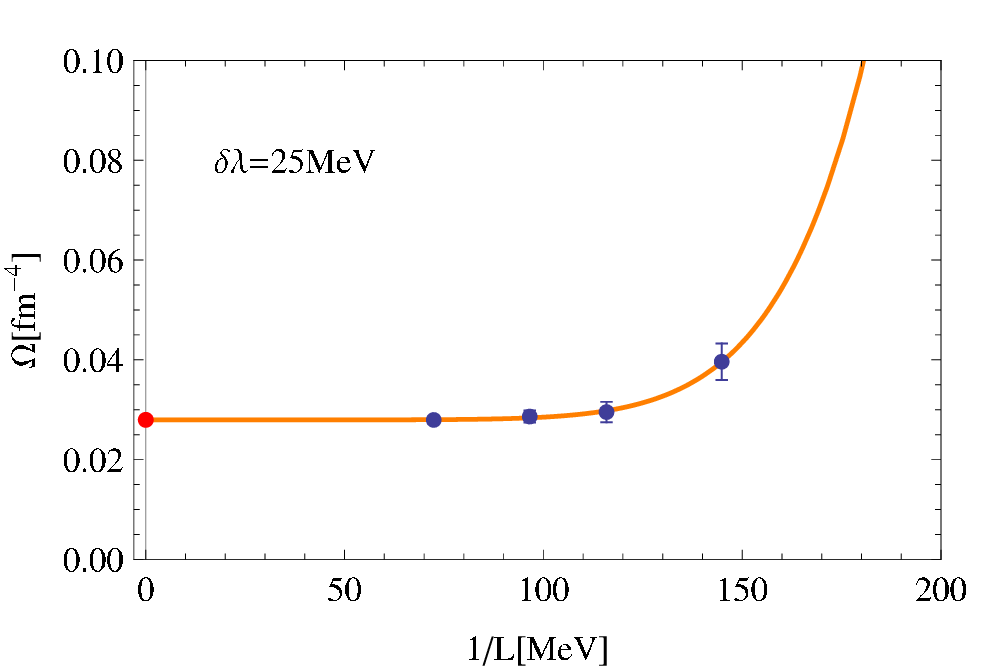}
     }
     \vskip 0.00in
     \caption{Top: behavior of $\cop_A(\lambda)$ and $\rho_{ch}(\lambda)$ in quenched QCD, 
     as discussed in the text. Bottom: infinite volume extrapolation of $\chps$ (left)
     and $\Omega$ of Sec.~{\bf 6} (right).}
    \label{fig:inf_volume}
    \vskip -0.45in
\end{center}
\end{figure} 

\medskip
\noindent{\bf 3. Chiral Polarization Scale.} In the following few sections we present
elements of evidence in support of the above statements. In all of our calculations we 
analyze the eigenmodes of overlap Dirac operator (parameters $r=1$, $\rho=26/19$) 
to ensure proper chirality properties at the regularized level.
The band of chirally polarized low--energy modes in quenched QCD was found in 
Ref.~\cite{Ale10A}. This work also showed that the point in the Dirac spectrum 
separating polarization from anti--polarization, namely $\chps$, is non--zero in 
the continuum limit at fixed physical volume. However, to be considered a true dynamical 
scale in the theory, $\chps$ needs to be demonstrably distinct from infrared cutoff.

To supply this crucial element, we calculated $\cop_A(\lambda)$ and $\rho_{ch}(\lambda)$ 
in quenched QCD at fixed lattice cutoff and several volumes. In particular, we work with 
Wilson lattice gauge theory at $\beta\!=\!6.054$ which translates into lattice spacing 
$a\!=\!0.085$ fm when invoking reference scale $r_0\!=\!0.5$ fm. 
Symmetric lattices of $N^4$ sites were used with $N=16,20,24,32$. Low--lying Dirac 
eigenmodes were computed on 100 equilibrated configurations for each system with 
$\cop_A(\lambda)$ and $\rho_{ch}(\lambda)$ evaluated using the modes in the strip 
$(\lambda - \delta\lambda/2,\lambda + \delta\lambda/2)$. We note that, in this regularized
case, $\lambda$ refers to magnitude of the overlap eigenvalue.
In Fig.\ref{fig:inf_volume} (top) we show the result for the largest lattice, displaying  
the advertised behavior, including a clearly defined $\chps$. The value of coarse graining 
parameter $\delta\lambda =25$ MeV is sufficiently small so that its variation induces only 
negligible changes in $\chps$.

The volume dependence of $\chps$ is shown on the lower left plot of 
Fig.\ref{fig:inf_volume}. The curvature of the data strongly suggests a positive infinite 
volume limit. Note that the infrared cutoff $1/L$ is also shown. Chi--Squares 
for fits of the form $\chps(1/L) = \chps(0) + b\, (1/L)^n$, with integer $n$, revealed 
a deep narrow minimum at $n=3$. This fitting function was then used for the infinite volume 
extrapolation shown. Thus, with Wilson lattice regularization at $a=0.085$ fm, our 
calculation gives $\chps \approx 160$ MeV in the infinite volume limit of quenched QCD. 
Using the continuum extrapolation of Ref.~\cite{Ale10A} as a rough guide for the size 
of cutoff effects, we estimate the continuum value to be $\chps \approx 150$ MeV in this
regularization.

\begin{figure}[t]
\begin{center}
    \centerline{
    \hskip 0.08in
    \includegraphics[width=8.7truecm,angle=0]{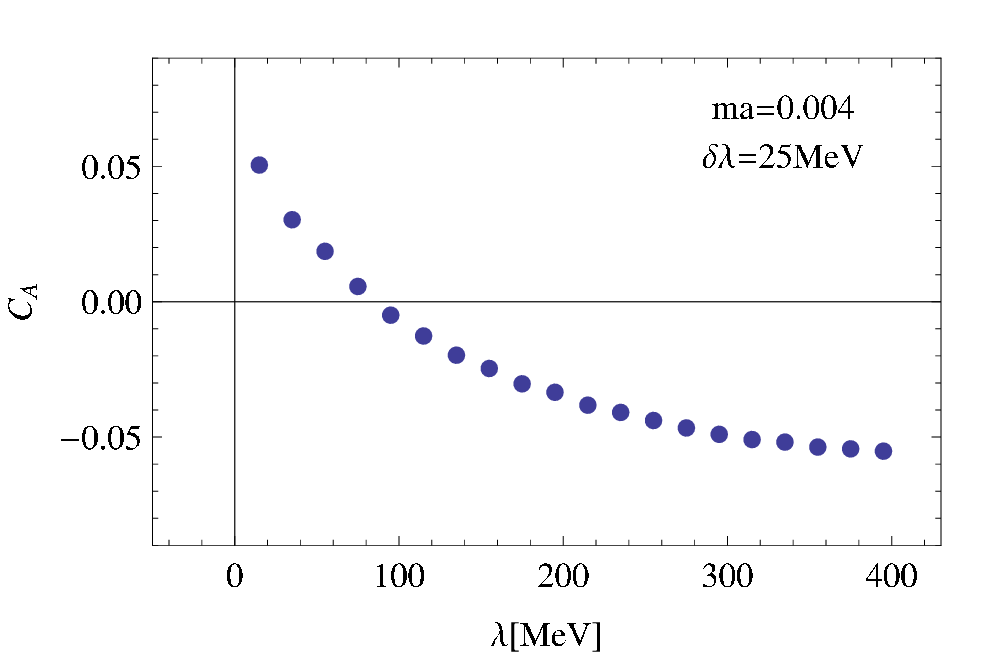}
    \hskip -0.10in
    \includegraphics[width=8.7truecm,angle=0]{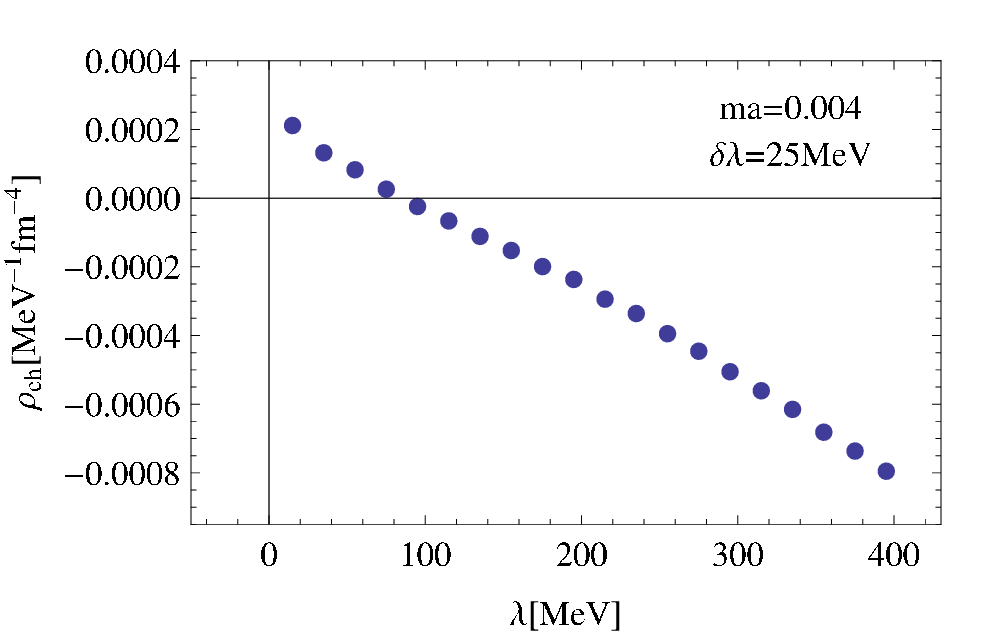}
     }
    \vskip -0.02in
    \centerline{
    \hskip 0.06in
    \includegraphics[width=8.7truecm,angle=0]{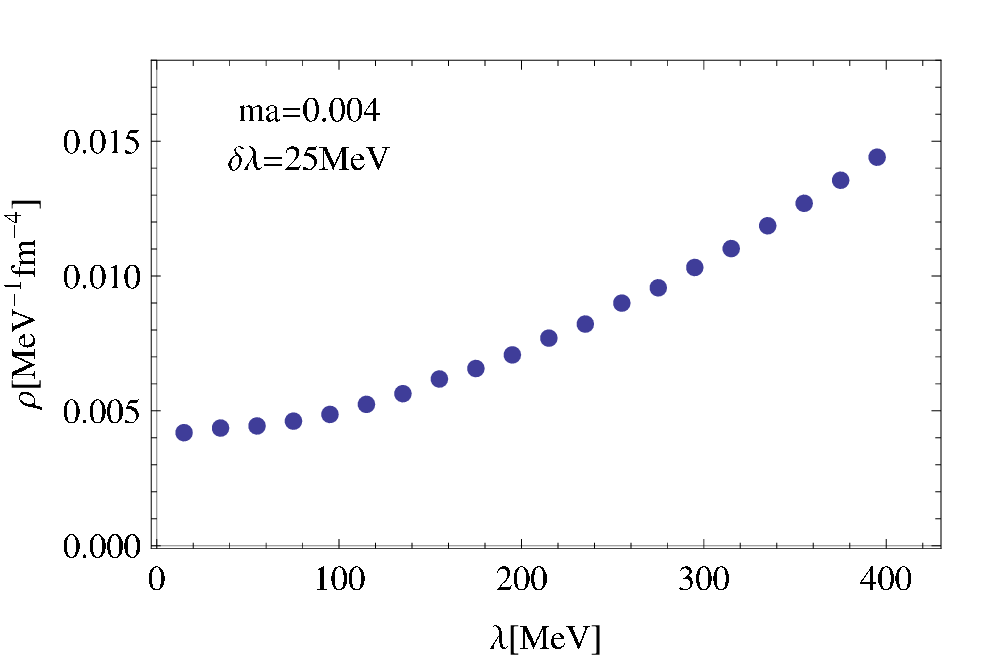}
    \hskip -0.10in
    \includegraphics[width=8.7truecm,angle=0]{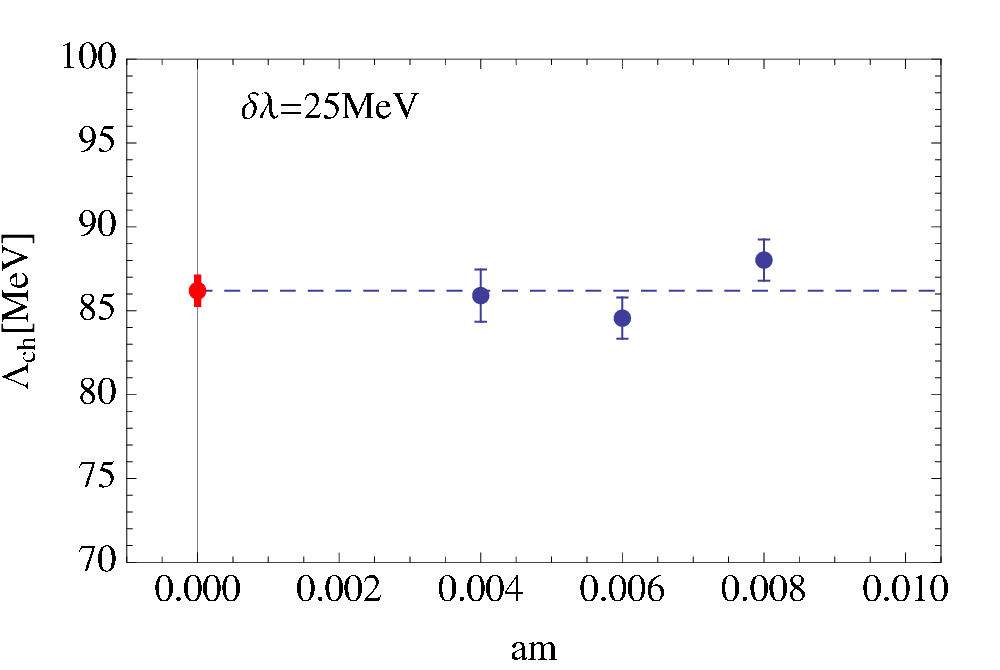}
     }
     \vskip 0.00in
     \caption{The behavior of $\cop_A(\lambda)$, $\rho_{ch}(\lambda)$ and $\rho(\lambda)$ 
     for the $m_l a = 0.004$ dynamical ensemble (see text). The light mass dependence
     of $\chps$ is shown in the lower right plot together with chiral extrapolation.}
    \label{fig:light_quarks}
    \vskip -0.45in
\end{center}
\end{figure} 

\vfill\eject

\noindent{\bf 4. Dynamical Light Quarks.} We now turn to the question whether light 
quarks in $N_f=$ 2+1 QCD could destroy chiral polarization of low--lying Dirac modes, 
thus disqualifying it as a dynamical feature of SChSB. Advantage of lattice 
regularizations respecting chiral symmetry is that such checks are meaningful at fixed 
cutoff. In this work we use the $32^3 \times 64$ domain wall (DW) ensembles generated by 
RBC/UKQCD collaborations as described in Ref.~\cite{RBC09A}. These three ensembles have fixed 
heavy bare mass $m_h a = 0.03$ and lattice spacing $a=0.085$ fm. The three light bare masses 
$m_l a = 0.008, 0.006, 0.004$ correspond to pion masses $m_\pi = 397, 350$ and $295$ MeV 
respectively. Since the scale has been set using the physical value of the $\Omega$
baryon, the heavy quark can be considered to have approximately the mass of the strange
quark.

We probe the vacua of the above DW lattice theories using the closely related overlap 
fermions. In particular, low--lying overlap Dirac eigenmodes were computed on 50 configurations 
from each ensemble and the characteristics of interest were evaluated. The functions 
$\cop_A(\lambda)$, $\rho_{ch}(\lambda)$ and $\rho(\lambda)$ for ensemble with lightest quarks 
are shown in Fig.~\ref{fig:light_quarks}. Note that the behavior consistent with 
{\bf Conjecture 1} is obtained, with the value of $\chps$ smaller relative to the quenched 
case, as expected. The important feature of our results is that the range of $m_l$ considered 
involves masses sufficiently small so that there is no appreciable change of $\chps$ detected.
This is shown on the lower right plot of Fig.~\ref{fig:light_quarks} together with 
extrapolation to chiral limit via constant fit. We estimate $\chps \approx 86$ MeV in 
the massless limit of this regularization, with the naive ($N_f$=0 based) expectation of 
the continuum value to be 5-10\% lower.

\begin{figure}[t]
\begin{center}
    \vskip -0.2in
    \centerline{
    \includegraphics[width=17.0truecm,angle=0]{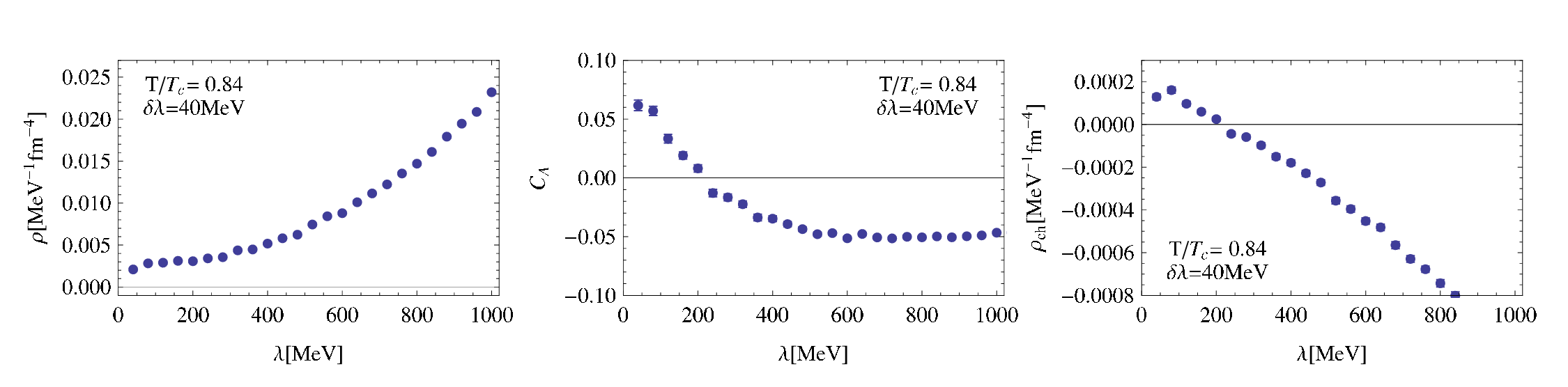}
     }
    \vskip -0.12in
    \centerline{
    \includegraphics[width=17.0truecm,angle=0]{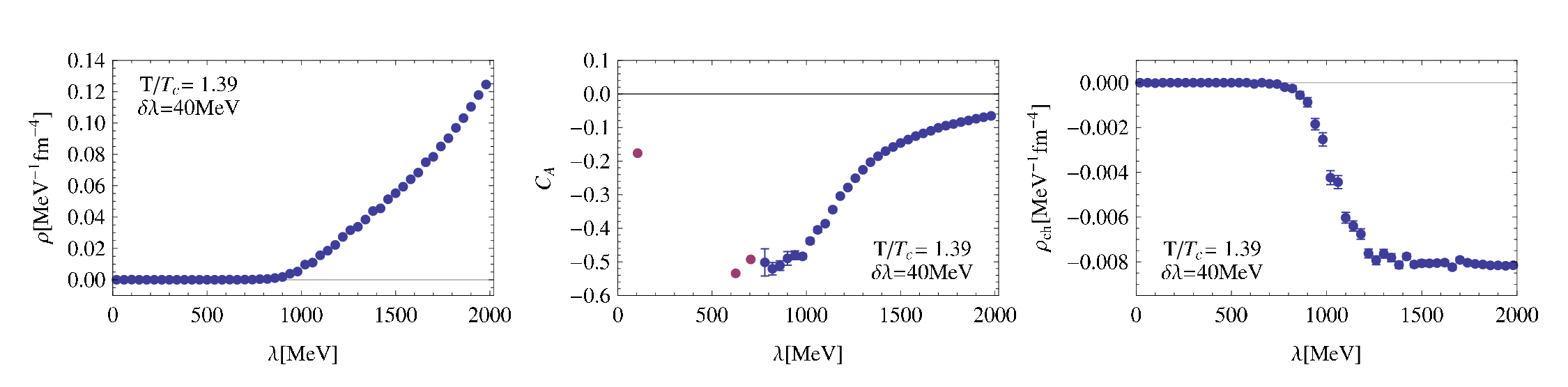}
     }
     \vskip -0.1in
     \caption{Top: functions $\rho(\lambda)$, $\cop_A(\lambda)$ and $\rho_{ch}(\lambda)$ 
     in quenched QCD at $T/T_c=0.84$, with 200 configurations. Bottom: same at $T/T_c=1.39$
     with 100 configurations.}
    \label{fig:fintmp_084_139}
    \vskip -0.45in
\end{center}
\end{figure} 

\medskip
\noindent{\bf 5. Finite Temperature.}
To check whether chiral polarization indeed goes hand in hand with mode condensation,
as proposed in {\bf Conjecture 2}, it is necessary to investigate systems in which the latter 
is expected to be absent. One way to do this is to turn on the temperature. For this purpose, 
we studied overlap Dirac eigenmodes on $20^3 \times N_t$ lattices of quenched QCD. The setup 
is otherwise identical to that described in Sec.~3. In particular, the same (Wilson) action 
is used at identical gauge coupling, corresponding to $a=0.085$ fm. 

Dirac mode condensation is expected to disappear in the vicinity of Polyakov line phase 
transition temperature $T_c$. Continuum--extrapolated value of $T_c$ is well 
known~\cite{Kar97A}, and in Fig.~\ref{fig:fintmp_084_139} we show the relevant eigenmode 
characteristics at $T/T_c=0.84$ ($N_t=10$), and at $T/T_c=1.39$ ($N_t=6$). 
Following the standard practice in quenched theory~\cite{Cha95A,Kov08A}, we compute 
fermionic observables over configurations from the ``real $Z_3$'' Polyakov line sector 
in order to facilitate smooth connection to simulations with dynamical quarks. 

As expected, the behavior of $\rho(\lambda)$ at $T/T_c=0.84$ is consistent with mode condensation 
while that at $T/T_c=1.39$ with its absence. The associated $\cop_A(\lambda)$ reflects this
with positive core of chiral polarization in the former case and the strictly negative
values in the latter. Note that the red points in the plot for $T/T_c=1.39$ signify bins with 
too few eigenmodes for an error bar to be estimated. It is feasible that, in the limit of 
infinite statistics, $\cop_A(\lambda)$ is defined everywhere, and there is no true spectral 
exclusion region. Chiral density $\rho_{ch}(\lambda)$ is of course not affected by sparse 
occurrence of low--lying eigenmodes and exhibits a behavior described by {\bf Conjecture 2}.  
Detailed account of these calculations, and of the correspondence, will be presented 
elsewhere~\cite{Ale12B}.

\medskip
\noindent{\bf 6. The Uses: Order Parameters and Finite Volume.} It is interesting to 
elaborate on some of the consequences stemming from the proposed connection of Dirac 
mode condensation to dynamical chiral polarization. To summarize the ingredients involved, 
consider the scalar fermionic density corresponding to valence quark mass $m_v$, namely 
\begin{equation}
   S(m_v,M) \equiv 
   \Big \langle\, \Trs\, ( D + m_v )^{-1}_{x,x} \, \Big \rangle_M   \,=\,
   \int_0^\infty d\lambda \, \frac{2 m_v}{m_v^2 + \lambda^2} \, \rho(\lambda, M)   
   \label{eq:710}
\end{equation}
Thus, the scalar density associated with dynamical flavor $f$ is 
$\langle \psibar\psi_f\rangle_M = -S(m_f,M)$. We say that the theory exhibits 
``valence'' SChSB, or broken valence chiral dynamics, if $\lim_{m_v \to 0} S(m_v,M) > 0$. 
Let's further define the associated chiral scalar density via the spectral relation
\begin{equation}
   S_{ch}(m_v,M) \equiv 
   \int_0^\infty d\lambda \, \frac{2 m_v}{m_v^2 + \lambda^2} \, \rho_{ch}(\lambda, M) 
   \label{eq:720}
\end{equation}
For both $S$ and $S_{ch}$ we have Banks--Casher type of relations, namely
\begin{equation}
   \lim_{m_v\to 0} S(m_v,M) = \pi \lim_{\lambda \to 0} \rho(\lambda,M)
   \qquad\quad
   \lim_{m_v\to 0} S_{ch}(m_v,M) = \pi \lim_{\lambda \to 0} \rho_{ch}(\lambda,M)
   \label{eq:730}
\end{equation}
with the former one expressing that Dirac mode condensation (at arbitrary $M$) is equivalent 
to valence SchSB. The equations above assume infinite volume. 

Conclusions of this paper imply that $S(0,M)$, $\rho(0,M)$, $S_{ch}(0,M)$,
$\rho_{ch}(0,M)$ and $\chps(M)$ can only be strictly positive simultaneously, thus
extending the standard relation involving just $S$ and $\rho$. 
The appearance of chiral polarization scale in the above mix is conceptually interesting since 
it transforms Dirac mode condensation (and SChSB) from strictly infrared ($\lambda=0$) effect 
into a feature that can be probed at scale $\chps$. This suggests that the behavior in systems 
of size $1/L \approx \chps$ should be sufficient to indicate whether given theory is  
condensing. Note that, unlike the ``$\lambda=0$'' quantifiers, the positivity of $\chps$ 
is compatible with finite volume. Indeed, to arbitrary function $\rho_{ch}(\lambda)$ on 
$[0,\infty)$ for which there exist $\lambda_1 < \lambda_2$ such that 
$\rho_{ch}(\lambda)\le 0$ on $(\lambda_1,\lambda_2)$, we assign $\chps \in [0,\infty)$ as 
the least $\lambda_1$ with that property. Chiral densities with $\chps$ assigned in this 
manner are positive on $[0,\chps)$ except for possible isolated zeros. Thus, even in finite 
volume, where $\rho_{ch}(0)=0$, scale $\chps$ can be strictly positive, e.g. when 
$\rho_{ch}(\lambda)>0$ on $(0,\chps)$. This was clearly seen in calculations presented 
in this work.

The above considerations make $\chps(M)$ an attractive ``order parameter'' for valence SChSB 
in SU(3) gauge theories, and for a physical SChSB in chiral corners of the theory space. 
Potentially useful alternative serving the same purpose is given by
\begin{equation}
  \Omega(M) \equiv \lim_{\Lambda \to \chps(M)^+} 
  \int_0^\Lambda d \lambda \, \rho_{ch}(\lambda,M) 
  \label{eq:740}
\end{equation}
i.e. total chiral polarization per unit volume. Note that $\Omega$ is a more general 
quantifier that would supersede $\chps$ in unlikely situations with chiral polarization 
entering only via a positive $\delta(\lambda)$ term in $\rho_{ch}(\lambda)$.  Indeed, 
this corresponds to $\chps=0$ but $\Omega > 0$. From practical standpoint it is important 
that $\Omega$ exhibits smaller finite volume effects than $\chps$, as shown 
in Fig.~\ref{fig:inf_volume} (bottom right).

\medskip
\noindent{\bf 7. The Uses: Broken Chiral Dynamics.} Assuming {\bf Conjectures 1,2} 
withstand further scrutiny, the dynamical information they carry offers additional 
possibilities. Here we point out that, since broken chiral dynamics is associated 
with chiral polarization while the symmetric behavior with anti--polarization, one 
may characterize the situation for massive quark by the balance of the two tendencies
in the associated scalar density. This is what $S_{ch}$ of Eq.~(\ref{eq:720}) has been 
designed for. Indeed, the integrands of Eqs.~(\ref{eq:710},\ref{eq:720}), 
namely contributions of modes at value $\lambda$, are identical up to a factor
$\cop_A$, since $\rho_{ch} = \rho \, \cop_A$. The meaning of $S_{ch}(m_v,M)$ is thus 
clarified by the fact that $S_{ch}(m_v,M)/S(m_v,M)$ expresses the average chiral 
polarization content of scalar density at valence mass $m_v$. If this average is 
negative in condensing theory, the dynamics of a probing massive quark has no 
resemblance to its polarized massless counterpart. On the other hand, if the average 
is positive, the dynamics of the quark in question is under the influence of broken 
(valence or dynamical) chiral symmetry. The two kinds of behavior are distinguished 
simply by the sign of $S_{ch}$. 

Elaborating on this, if the theory doesn't condense then $\rho_{ch}(\lambda) \le 0$, 
and $S_{ch}(m_v)\le 0$ for arbitrary $m_v$, consistently with chiral symmetry 
(valence or dynamical) not being broken. But in the condensing case 
$\rho_{ch}(\lambda) > 0$ on $[0,\chps)$, implying that $S_{ch}$ is positive for 
sufficiently small $m_v$. Indeed, the factor $m_v/(m_v^2 + \lambda^2)$ plays role 
analogous to low--pass filter with passing spectral range controlled by $m_v$. 
In fact, there is $m_{ch}>0$ with $S_{ch}(m_{ch})=0$, separating the ``light quark'' 
region driven mainly by broken chiral dynamics from the ``heavy quark'' region where 
chiral considerations play little role. 
The consequence of this for dynamical quark $m_f \in M$ is obvious: if $m_f < m_{ch}(M)$ 
then it is in the broken regime. Removing the reference to valence quarks, the above 
translates into the statement 
\begin{equation}
   \mbox{\rm flavor} \; m_f \in M \;\, \mbox{\rm broken chiral}  
   \qquad \Longleftrightarrow \qquad S_{ch}(m_f,M) >0
   \label{eq:810}
\end{equation}
Thus, the insight gained from the proposed association of SChSB with chiral polarization 
may provide a microscopic basis to judge the applicability of chiral perturbation 
theory. In particular, for $M=(m_u,m_d,m_s)$ of ``real-world'' QCD, 
one can inquire whether $S_{ch}(m_s,M)>0$. 

It should be noted that densities $S$ and $S_{ch}$ involve a power cutoff divergence 
at non--zero quark mass~\cite{Her99A}, which needs to be subtracted at the 
{\em spectral level} for the above considerations to become meaningful. Discussion in 
this section thus implicitly assumed that the $\lambda$--dependent subtraction of mode 
density $\rho(\lambda) \rightarrow \rho_{sub}(\lambda)$, needed for proper spectral 
definition of $S$, is also used in the definition of $S_{ch}$. Related details will be 
discussed elsewhere~\cite{Ale12C}. 

\begin{figure}[t]
\begin{center}
    \centerline{
    \hskip 0.00in
    \includegraphics[width=10.0truecm,angle=0]{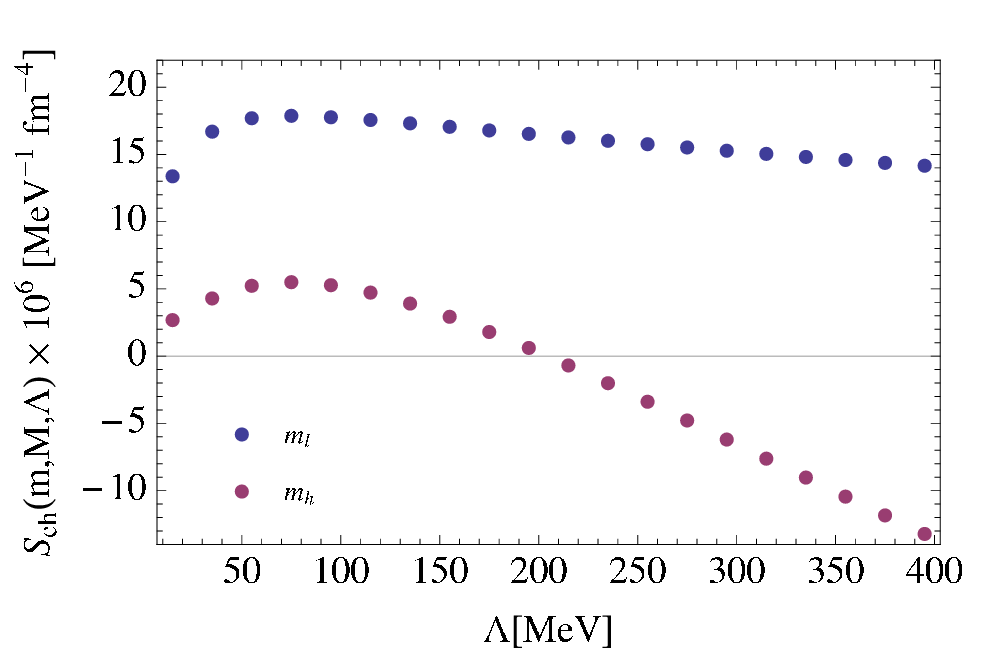}
     }
     \vskip -0.10in
     \caption{Cumulative chiral scalar density for ``light'' and ``heavy'' quark of $N_f=$2+1
     QCD. See explanation in the text.}
     \label{fig:strange}
    \vskip -0.45in
\end{center}
\end{figure} 

In order to see the characteristic behavior, as well as to obtain the first estimate 
for the sign of $S_{ch}(m_s,M)$ in QCD, we present the result of a simplified calculation
using the $m_l a=0.004$ dynamical ensemble of Sec.~4. Here the pion mass is still about
twice the physical value, but the heavy quark $m_h a=0.03$ is at the strange mass. 
In Fig.~\ref{fig:strange} we show the plots of cumulative chiral scalar density
$S_{ch}(m,M,\Lambda)$ defined by Eq.~(\ref{eq:720}) via replacement 
$\int_0^\infty d\lambda \rightarrow \int_0^\Lambda d\lambda$. Note that both light 
$m \leftrightarrow m_l$ and heavy $m \leftrightarrow m_h$ cases are shown. Since 
overlap Dirac probe is used on domain--wall sea, the bare quark masses in the former 
need to be appropriately matched. In case of light quark we used the results of 
Ref.~\cite{Luj12A} for match based on the pion mass. The heavy mass was scaled up 
proportionally, which is an approximation. The $\lambda$--dependent subtraction in 
$\rho_{ch}$ was ignored but its effect is expected to be small at low--energy scales 
involved here. 

As one can see, the behaviors for light and heavy quark are quite different. In the 
light case there is a very slow decrease of $S_{ch}(\Lambda)$ that will at larger scales  
become further moderated by the subtraction, and the positive $\Lambda \to \infty$ 
limit will presumably result. On the other hand, the density for the ``strange quark'' 
changes sign already at around $\Lambda$= 200 MeV, and the negativity appears rather
robust even in light of the approximations involved. 

\medskip
\noindent{\bf 8. The uses: Finite Temperature.} 
The features of chiral symmetry discussed here may be of particular interest for studying 
QCD at finite temperature. Indeed, the characterization of changes in chiral properties 
of heated quarks and gluons is usually limited to monitoring of scalar densities 
$-\langle \psibar\psi_f \rangle_{M,T}=S(m_f,M,T)$ (or their susceptibility) 
as a function of temperature. However at non--zero quark masses, such as those of 
real world, chiral symmetry is explicitly broken and trends in $S(m_f,M,T)$ may not be 
most reflective of changes in the dynamics of the quark--gluon system. After all, 
crossover is currently a preferred scenario at physical quark masses~\cite{Aok06A}, 
without a unique point separating definite qualitative behaviors.

However, if one takes a ``dynamical mechanism'' viewpoint, instead of 
the ``thermodynamic'' one, then there could be definite chirality--related properties 
that qualitatively change due to temperature. In the context proposed here, it is first of 
all natural to distinguish when the system supports density of chirally polarized quark 
modes. This feature is adjudicated by the positivity of $\chps(M,T)$, and is equivalent 
to Dirac mode condensation as well as valence chiral symmetry breakdown. For arbitrary 
$M$ with $\chps(M,T=0)>0$, we expect the existence of temperature $T_{ch}(M)>0$ such that
\begin{equation}
   \chps(M,T)\; 
   \begin{cases} 
   \;> 0 & \mbox{for} \quad T < T_{ch}(M) \\ 
   \;= 0 & \mbox{for} \quad T > T_{ch}(M) 
   \end{cases}
   \label{eq:910}
\end{equation}
Thus, $T_{ch}>0$ should exist in N$_f$=2+1 QCD at arbitrary quark masses. We emphasize 
that $T_{ch}$ really characterizes the {\em gluonic part} in the dynamics of the massive 
quark--gluon system. Indeed, at $T>T_{ch}$ the relevant gauge backgrounds no longer support 
the feature (whatever it may be) allowing for chirally polarized modes to exist, and to 
facilitate broken dynamics of massless valence probe. When dynamical quarks are heavy, such 
change may not be crucial for their own dynamics, but this becomes so when they are light. 
Moreover, given the smooth approach to chiral limit seen in Fig.~\ref{fig:light_quarks}, 
$T_{ch}(M)$ probably marks the disappearance of the sought--after dynamical feature in glue, 
producing SChSB. Note also that, away from chiral limit, $\Lambda_{ch}(M,T)$ is a strictly 
non--local ``order parameter'' in terms of dynamical quark and gluon fields. 
Thus, its non--analytic nature at {\em finite} $T_{ch}(M)$ would not contradict 
the observed crossover behavior of standard thermal observables at physical 
point~\cite{Aok06A}.

What about the characteristics relevant directly to dynamical quarks? In zero temperature 
QCD, chiral issues are mainly important because it is believed that ``u'' and ``d'' 
quarks are driven by broken chiral dynamics. It is then natural to expect that if given 
flavor(s) becomes no longer broken due to thermal agitation, the physical properties 
in the heated quark--gluon system qualitatively change. The degree of 
``broken quark dynamics'' is controlled by $S_{ch}(m_f,M,T)$ in a manner 
specified by Eq.~(\ref{eq:810}). For each flavor $f$ with $S_{ch}(m_f,M,T=0)>0$ we 
expect the existence of the temperature $T_{ch,f}(M)$ such that 
\begin{equation}
  S_{ch}(m_f, M,T)\; 
  \begin{cases} 
  \;> 0  & \mbox{for} \quad T < T_{ch,f}(M) \\ 
  \;< 0  & \mbox{for} \quad T > T_{ch,f}(M) 
  \end{cases}
  \label{eq:920}
\end{equation}

In summary, if the number of broken flavors at zero temperature is $0 < N_{f,b} \le N_f$, then 
there are $N_{f,b}+1$ transition temperatures characterizing chiral behavior, satisfying 
\begin{equation}
   T_{ch} \,\ge\,  T_{ch,1} \,\ge\, T_{ch,2} \,\ge\, \ldots \,\ge\, T_{ch,N_{f,b}}
   \label{eq:930} 
\end{equation}
where flavors are assumed to be ordered by mass $m_1 \le m_2 \le ...\le m_{N_f}$. 
The first relation becomes equality only when $m_1=0$ while the rest only when 
the corresponding pair of quarks is mass--degenerate. We emphasize again that these 
transitions may not correspond to phase transformations in the usual thermodynamic sense. 
Rather, they reflect qualitative changes in the dynamical behavior of participating quarks 
and gluons.
In this way, the insight of chiral polarization offers a different perspective on QCD 
at finite temperature. 

\vfill\eject

\noindent{\bf Acknowledgments:} 
We are indebted to RBC/UKQCD collaborations for sharing their dynamical configurations,
and to Mike Lujan/$\chi$QCD collaboration for providing corresponding overlap eigenmodes.
Thanks to Mike Creutz, Tam\'as Kov\'acs and Stefan Sint for communications related to this 
project, and to Mingyang Sun for help with graphics. 
A. A. is supported in part by U.S. Department of Energy under the grant 
DE-FG02-95ER-40907 and by U.S. National Science Foundation under CAREER grant PHY-1151648.
I.H. acknowledges the support by Department of Anesthesiology at the University
of Kentucky.

\end{document}
\bye